\documentclass[journal]{IEEEtran}

\usepackage{amsmath,amssymb,amsfonts}
\usepackage{amsthm}
\usepackage{cite}
\usepackage[pdftex]{graphicx}
\usepackage[caption=false,font=footnotesize]{subfig}
\usepackage{verbatim}
\usepackage{array}
\usepackage{algorithm}
\usepackage{stfloats}
\usepackage{url}
\usepackage{caption}
\usepackage{color,soul}
\usepackage[framemethod=tikz]{mdframed}
\usepackage{xcolor}

\begin{document}

\title{The Internet of Bio-Nano Things in Blood Vessels:\\ System Design and Prototypes}

\author{Changmin~Lee,~\IEEEmembership{Student~Member,~IEEE,}
	Bon-Hong~Koo,~\IEEEmembership{Student~Member,~IEEE,}\\~
	Chan-Byoung~Chae,~\IEEEmembership{Fellow,~IEEE}, and~Robert Schober,~\IEEEmembership{Fellow,~IEEE}
	\thanks{C. Lee, B.-H. Koo, and C.-B. Chae are with the School of Integrated Technology, Yonsei University, Seoul, Korea (e-mail: cm.lee@yonsei.ac.kr; harpeng7675@yonsei.ac.kr; cbchae@yonsei.ac.kr). R. Schober is with Institute of Digital Communications, Friedrich-Alexander University of Erlangen-Nuremberg, Erlangen, Germany (email: robert.schober@fau.de).}}
\maketitle

\begin{abstract}
In this paper, we investigate  the Internet of Bio-Nano Things (IoBNT) which relates to networks formed by molecular communications. By providing a means of communication through the ubiquitously connected blood vessels (arteries, veins, and capillaries), molecular communication-based IoBNT enables a host of new eHealth applications. For example, an organ monitoring sensor can transfer internal body signals through the IoBNT for health monitoring applications. We empirically show that blood vessel channels introduce a new set of challenges for the design of molecular communication systems in comparison to free-space channels. We then propose cylindrical duct channel models and discuss the corresponding system designs conforming to the channel characteristics. Furthermore, based on prototype implementations, we confirm that molecular communication techniques can be utilized for composing the IoBNT. We believe that the promising results presented in this work, together with the rich research challenges that lie ahead, are strong indicators that IoBNT with molecular communications can drive novel applications for emerging eHealth systems.
\end{abstract}


\section{Introduction}
The increasing global population has led to a mounting set of challenges. Significant social issues may require solutions from an entirely new perspective; for example, an aging society will have a high demand for an advanced healthcare system. While the acceleration of medical research is indispensable for overcoming diseases, bio-nano engineering can assist in the goal of preventing advanced pathogenesis. In particular, precautionary actions can play a pivotal role in the wellbeing of older people, who may have lower self-healing ability. We expect that the Internet of Bio-Nano Things (IoBNT) may allow preventative monitoring on a daily or even hourly basis, in contrast to conventional health monitoring which is currently limited to annual or monthly examinations~\cite{Pramanik2020, Zafar2021}. It is worth considering the role of the IoBNT in health monitoring systems, which we refer to as eHealth.

We spotlight health monitoring systems enabled by bio-nano machines which function and communicate with each other inside a body. Such systems require substantial provisions for biocompatibility and energy efficiency. The communication function of conventional robots for eHealth mainly relies on electromagnetic (EM) waves; however, this approach poses several problems, such as the development of nanoscale \mbox{actuators}, antennas, or body absorption of tera-Hertz band \mbox{frequency}. In this work, we introduce molecular communications for eHealth as an alternative technology.

Molecular communication is a bio-inspired technology wherein components of the system exchange information via molecules. For example, in nature, an ant colony \mbox{implements} swarm intelligence using pheromone signals, and organs work in harmony by regulating each other through \mbox{hormones}. Studies have shown that molecular communications is \mbox{advantageous} over radio frequency (RF) technology~\cite{Akyildiz2015, Akyildiz2019, Guo2016, Farsad2016} in certain applications. In one study~\cite{Akyildiz2015}, the authors claimed that synthetic biology- and nanotechnology-based technologies could be used to construct bio-nano units for \emph{in-vivo} applications. They noted that EM radiation can have negative effects on the body. A consequential prototype study by the same authors was reported in~\cite{Akyildiz2019}. W. Guo, et al.~\cite{Guo2016} compared the channel characteristics of molecular and RF communications. Numerous studies on molecular communications for general scenarios have been conducted in~\cite{Farsad2016}.

In this work, we investigated health monitoring systems comprised of scattered bio-nano things inside blood vessels. The system exploits blood vessels as molecular communication channels. This requires a distinct channel model, whereas most existing studies on molecular communications have focused on free-space diffusive channels. Spatial restrictions and the drift present in blood vessels are the main differences compared to free-space diffusive channels. The spatial limitations can be approximated using a cylindrical duct shape. There have been studies on such cases~\cite{Jamali2019, Lo2019, Schafer2021, Dhok2022, Dhok2022a, Varshney2018, Zoofaghari2019, Arjmandi2021, Lee2020, Koo2020}, where the authors in~\cite{Jamali2019} introduced various channel models with varying velocity profiles and shape parameters. In particular, the channel model for a special velocity profile, Poiseuille flow, was introduced in~\cite{Lo2019, Schafer2021}. The authors of~\cite{Varshney2018, Zoofaghari2019, Arjmandi2021} suggested a channel model considering biological cylindrical environment characteristics such as the molecule degradation effect and arbitrary boundary. In contrast, testbed realizations of meso- and nano-scale duct channels were reported in~\cite{Lee2020, Koo2020}. 

The remainder of this paper is organized as follows. In Section~\ref{health_monitoring_channel}, we introduce IoBNT based on blood vessel networks, describe an eHealth system, and specify channel models for the proposed molecular communication system. In Section~\ref{system_design}, we describe the concrete design and requirements of the health monitoring systems. In Section~\ref{testbed}, we present a testbed modeling channel environments similar to the human blood vessel system. Finally, in Section~\ref{conclusion}, we conclude our paper and address open challenges for advanced health monitoring.

\begin{figure*}[!t]
	\centering
	\includegraphics[width=1.5\columnwidth,keepaspectratio]{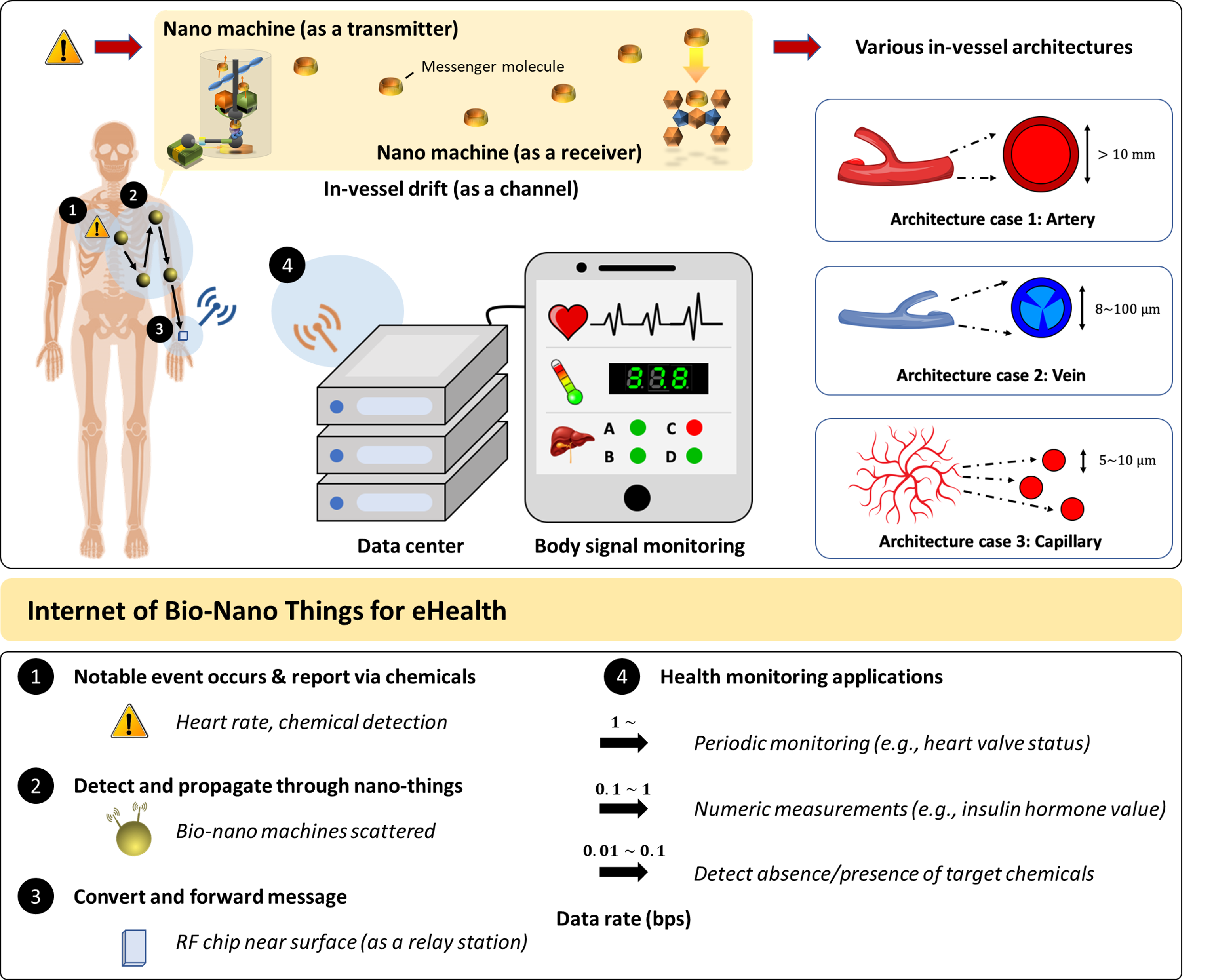}
	\caption{A conceptual depiction of the IoBNT for eHealth. The health monitoring scenario we spotlight here employs nano-machines which can communicate with each other via molecules so that the system can have better biocompatibility than a system employing EM components. The machines are capable of adaptively transmitting and receiving target molecules in in-vessel environments with the help of state-of-the-art bio-nano machine technology. }
	\label{fig_monitoring_system}
\end{figure*}

\section{IoBNT \& Blood Vessel Environments}
\label{health_monitoring_channel}

\begin{figure*}[!t]
	\centering
	\includegraphics[width=1.5\columnwidth,keepaspectratio]{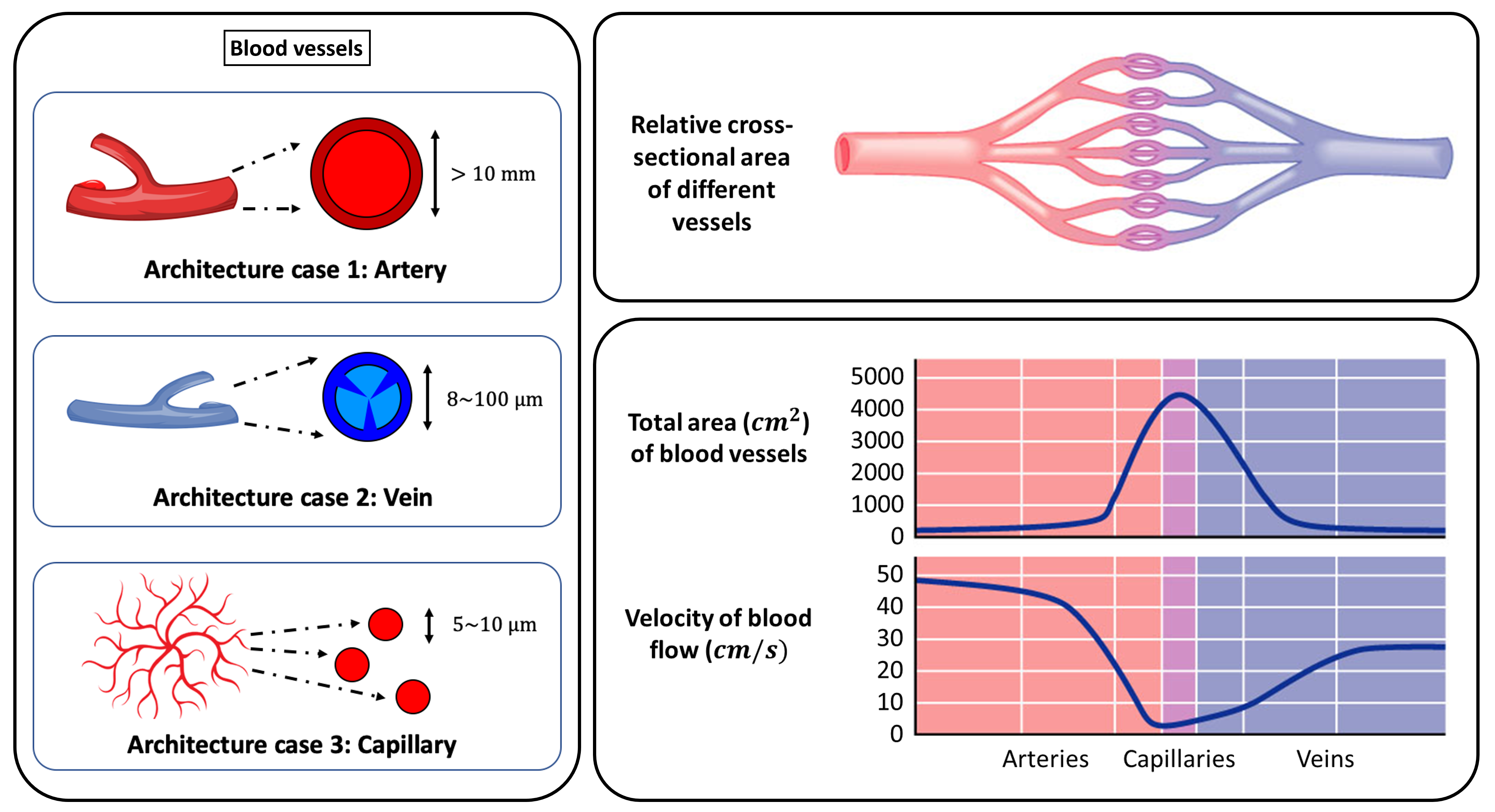}
	\caption{Characteristics of blood vessels (artery, vein, and capillary): the size of the cross-section, the total area of blood vessels, and the velocity of blood flow~\cite{Campbell2016}.}
	\label{fig_blood_vessels}
\end{figure*}

In this paper, we investigate the IoBNT, which utilizes molecular communications, especially in blood vessels. Blood vessels spread throughout the entire body and are good channels for building IoBNT, which can gather information at various locations in the body. Fig.~\ref{fig_monitoring_system} illustrates the conceptual depiction of an IoBNT-based health monitoring system comprising four main steps. In the first step, the sensors measure human body signals at various locations inside the body. These sensors should be harmless and operate under tight energy constraints. As some previous work suggested~\cite{Akyildiz2015, Abbasi2017, Akyildiz2019, Akyildiz2020}, nano- and bio-technology will form the basis for various bio-nano-based sensing systems. Second, bio-nano things deployed inside the body create a network through molecular communication via blood vessels and convey the measured signals from the sensors~\cite{Varshney2018, Misra2020, El-Fatyany2020, Dissanayake2021, Al-Zubi2022, Juwono2021}. Third, the platform developed in~\cite{Swaminathan2017} can be utilized when connecting the IoBNT network to the relay chip located under the skin. The system works based on galvanic coupling and enables communication through tissue. Finally, the chip is connected to electronic devices outside the body through RF technology such as Bluetooth. We can then exploit the collected body signals for medical applications.

For example, patients with cardiac disorders may use a continuous monitoring system after heart surgery. The sensor positioned near the heart measures cardiac signals, such as heartbeat rhythm. The molecular communication network forwards the cardiac signals, and users can monitor the function of the heart through a smartphone. Similarly, patients with diabetes may use the system to periodically monitor their blood sugar levels. Continuous monitoring of biological signals from the body can assist doctors in their diagnosis and lead to better prescription and/or treatment for previously difficult to capture symptoms.

From a system requirement perspective, such a communication system must be highly reliable and harmless to the patient. However, these systems do not necessarily require high transmission speeds, as the data are kept simple in most application scenarios. Specific components of IoBNT systems, such as transceivers and sensors, have been comprehensively investigated in previous studies~\cite{Akyildiz2015, Chude-Okonkwo2016, Akyildiz2019, Akyildiz2020, Yang2020}.

Many researchers have proposed blood vessel networks as channels for IoBNT networks in the human body. The blood vessels in the body can play a key role to compose the communication system that can cover the whole body. However, researchers mainly focused on the communication system itself without considering the effect of blood vessels~\cite{Varshney2018, Misra2020, El-Fatyany2020, Dissanayake2021, Al-Zubi2022, Juwono2021}. Therefore, we mainly deal with the communication system in the blood vessels.

The human blood vessel network is composed of three different types: arteries, veins, and capillaries, all of which have different channel characteristics. Therefore, we need to build an IoBNT network by considering the channel characteristics of blood vessels. Fig.~\ref{fig_blood_vessels} illustrates the characteristics of blood vessels in the human body~\cite{Campbell2016}.

The arteries have a much larger cross-section size than other types of blood vessels, as well as the fastest rate of blood flow. As a result, blood passing through arteries generally displays features of turbulent flow. In particular, the arteries close to the heart exhibit strong turbulent flow due to heartbeat-induced fluctuations in blood pressure. Naturally, blood pressure becomes more stable for arteries located farther away from the heart.

The veins have various cross-sectional sizes and flow speeds. Therefore, we should consider that the flow features vary according to the environment. The Reynolds number given by the radius of the vessel and the flow speed is an appropriate criterion for determining whether the flow is turbulent or laminar. On the other hand, veins often have negative blood pressure as they are located far from the heart. Therefore, biological valves assist in blood flow in veins.

The capillaries had the smallest cross-sectional sizes and flow speeds. Therefore, they usually exhibit laminar flow characteristics. Unusually, the capillary surface is permeable, so that nutrients can be conveyed to adjacent cells. The channel model differs from other blood vessels at this point.

The three blood vessels show different schemes owing to the above characteristics. Therefore, we should consider the optimal communication systems for each environment. However, channel studies of molecular communication for free space are difficult to apply to this system. The boundary, such as blood vessel walls, makes the movement of the molecules different. In addition, various blood vessel characteristics influence the channel. Therefore, we introduce a method for handling each factor to build communication systems and suggest system designs for IoBNT in the following section.

\begin{figure*}[!t]
	\centering
	\includegraphics[width=1.8\columnwidth,keepaspectratio]{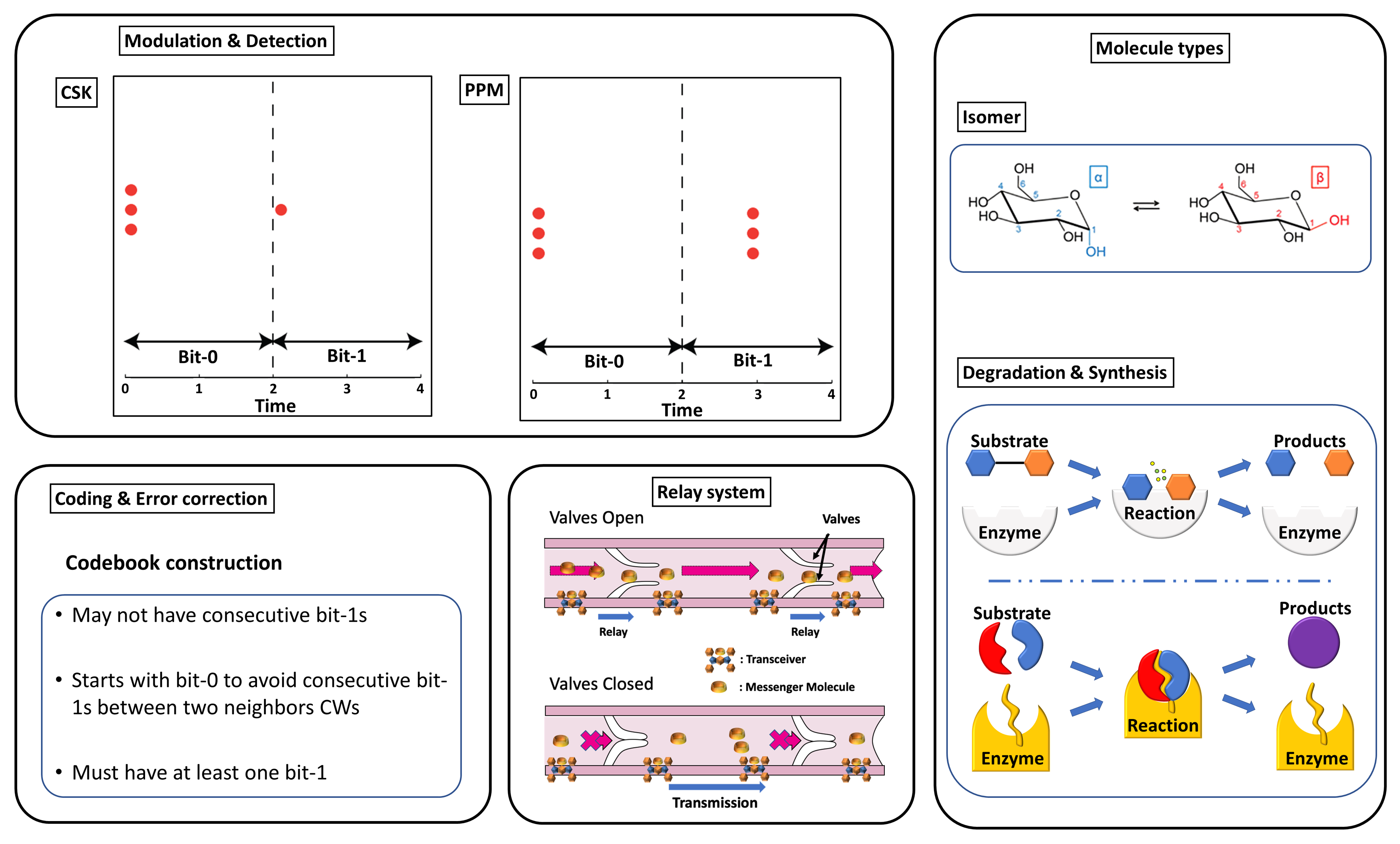}
	\caption{Proposed system design for the molecular communication in blood vessels.}
	\label{Architectures}
\end{figure*}


\section{System Design}
\label{system_design}
In this section, we investigate the system design of molecular communication systems for blood vessel channels for eHealth applications, as shown in Fig.~\ref{Architectures}.

The messenger molecules type, flow speed, cross-section size, and degradation existence should influence channel model development. Also, the positions of the transmitter and receiver should influence the channel model. We expect the transmitter and the receiver to be located at the surface because extra energy is required when the transceiver is located in inner blood vessels to keep the position in a realistic application view. Finally, we discuss the optimal system: modulation, detection, coding, error correction, and relay system based on each blood vessel channel model.

\subsection{Channel}
\begin{figure*}[!t]
	\centering
	\includegraphics[width=1.6\columnwidth,keepaspectratio]{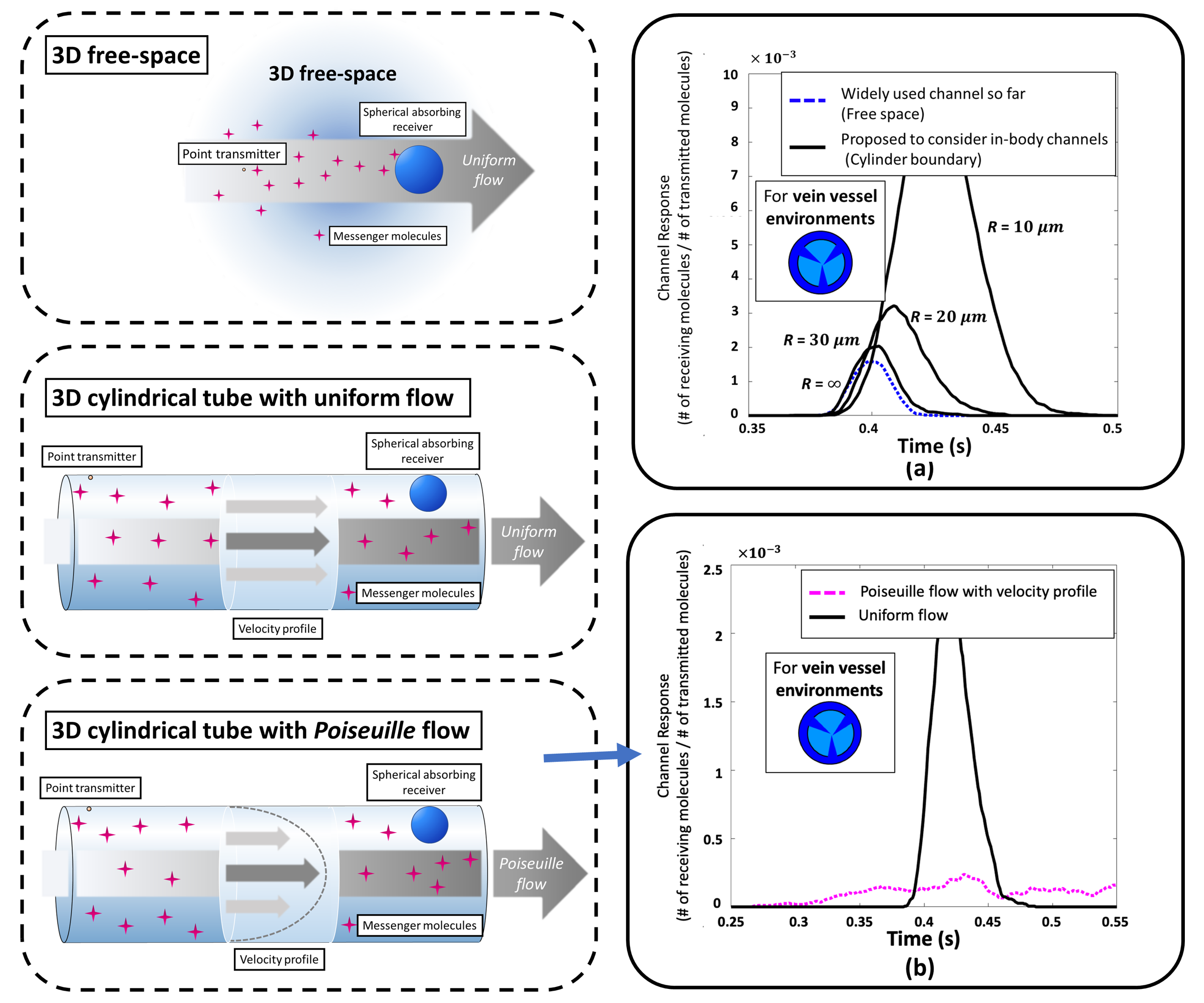}
	\caption{Illustration of channel environments (left) and channel responses in a vein blood vessel (right). In each channel environment (left), a point transmitter emits molecules and a spherical absorbing receiver counts the received molecules. The transmitter and receiver are assumed to adhere to the surface of the 3D cylindrical tube. Each environment under consideration has different flow phenomena by velocity profiles. For the channel responses (right), we propose adopting the cylindrical tube channel as in-body network channel models.}
	\label{fig_Channel_model}
\end{figure*}

Although they exhibit different characteristics, all types of blood vessels can be modeled as cylindrical ducts~\cite{Jamali2019, Lo2019, Schafer2021, Dhok2022, Dhok2022a}. Fig.~\ref{fig_Channel_model} illustrates three different channel models considered in the previous studies: the free-space channel, the cylindrical duct channel with uniform flow, and the cylindrical duct with Poiseuille flow. We only assume communication systems in blood vessels that show laminar flow, such as uniform flow and Poiseuille flow. Because it is difficult to handle communication systems in a turbulent flow.

We assume that the point transmitter emits the messenger molecules as carriers, and the spherical receiver absorbs them. The number of molecules captured at the receiver constitutes the channel response of molecular communications. For the cylindrical duct channel models, the transmitter is located at one end of the duct and adheres to the surface, with the receiver located at the other end of the duct on the surface. The channel responses illustrated in Fig.~\ref{fig_Channel_model} were obtained through computer simulations by applying the vein channel parameters.\footnote{The diffusion coefficient and the uniform flow velocity are 670~$\mu$$m^2/s$ and 0.5~$cm/s$, respectively, and the length of the cylindrical duct, the radius of the duct, and the receiver size are 2000~$\mu$$m$, 30~$\mu$$m$, and 5~$\mu$$m$, respectively.} 

Fig.~\ref{fig_Channel_model} illustrates the difference between the channel responses of the cylindrical duct channel and the free-space channel. The cylindrical duct channel yields a higher signal amplitude than the free-space channel, which indicates that the channel is more directional. However, the tail, which increases the inter-symbol interference (ISI), increases. As the radius of the duct increases, the cylindrical duct channel response becomes similar to the free-space channel response.

Furthermore, the effects of the different flow velocity profiles on the radial axis are shown in Fig.~\ref{fig_Channel_model}b. Poiseuille flow has a non-uniform velocity profile that depends on the liquid characteristics, flow speed, and size of the duct cross-section. The friction force between the tube surface and the liquid in the duct causes a non-uniform velocity profile. Typically, in small ducts such as blood vessels, this friction force cannot be ignored. The Péclet number, defined as the ratio of the rate of advection of a physical quantity caused by the flow to the rate of diffusion of the same quantity driven by an appropriate gradient (i.e., $Pe = \frac{Rv}{D}$, where $R$ is the radius of the tube, $v$ is the flow velocity, and $D$ is the diffusion coefficient.) can be used to describe the relative importance of flow and diffusion. A flow acts as a Poiseuille flow when the dispersion factor, $\alpha_d$ (${\alpha}_d = \frac{L}{Pe~R}$),  is smaller \mbox{than one}. This is because there is insufficient time to achieve uniform flow by radial diffusion if the dispersion factor is smaller than one. The first arrival of molecules is faster in the Poiseuille flow, and the tail of the channel response becomes longer in some cases.

Unlike prior work focusing on the free-space channel, in this work, we suggest utilizing cylindrical duct channel models for molecular communications in blood vessels. Generally, large blood vessels have uniform flow~\cite{Jamali2019, Dhok2022, Dhok2022a}, while others have Poiseuille flow, which exhibits a more complicated channel response~\cite{Lo2019, Schafer2021}. However, other blood vessel characteristics also influence flow conditions; therefore, we should select a proper flow channel model based on the Péclet number and dispersion factor calculated by the given blood vessel parameters.

\begin{figure*}[!t]
	\centering
	\subfloat[]{\includegraphics[width=1.6\columnwidth,keepaspectratio]{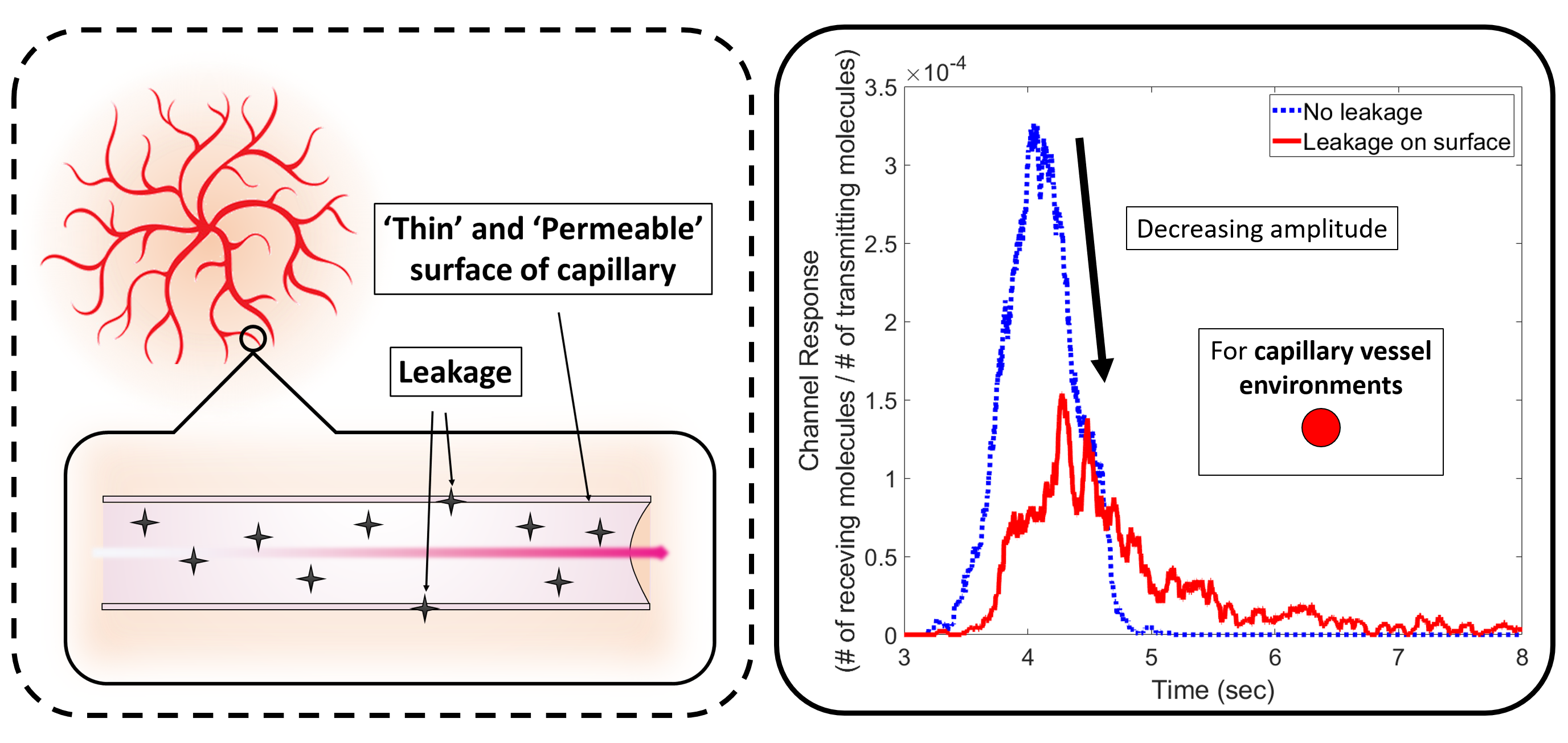}
		\label{fig_leakage}}
	\hfill
	\subfloat[]{\includegraphics[width=1.6\columnwidth,keepaspectratio]{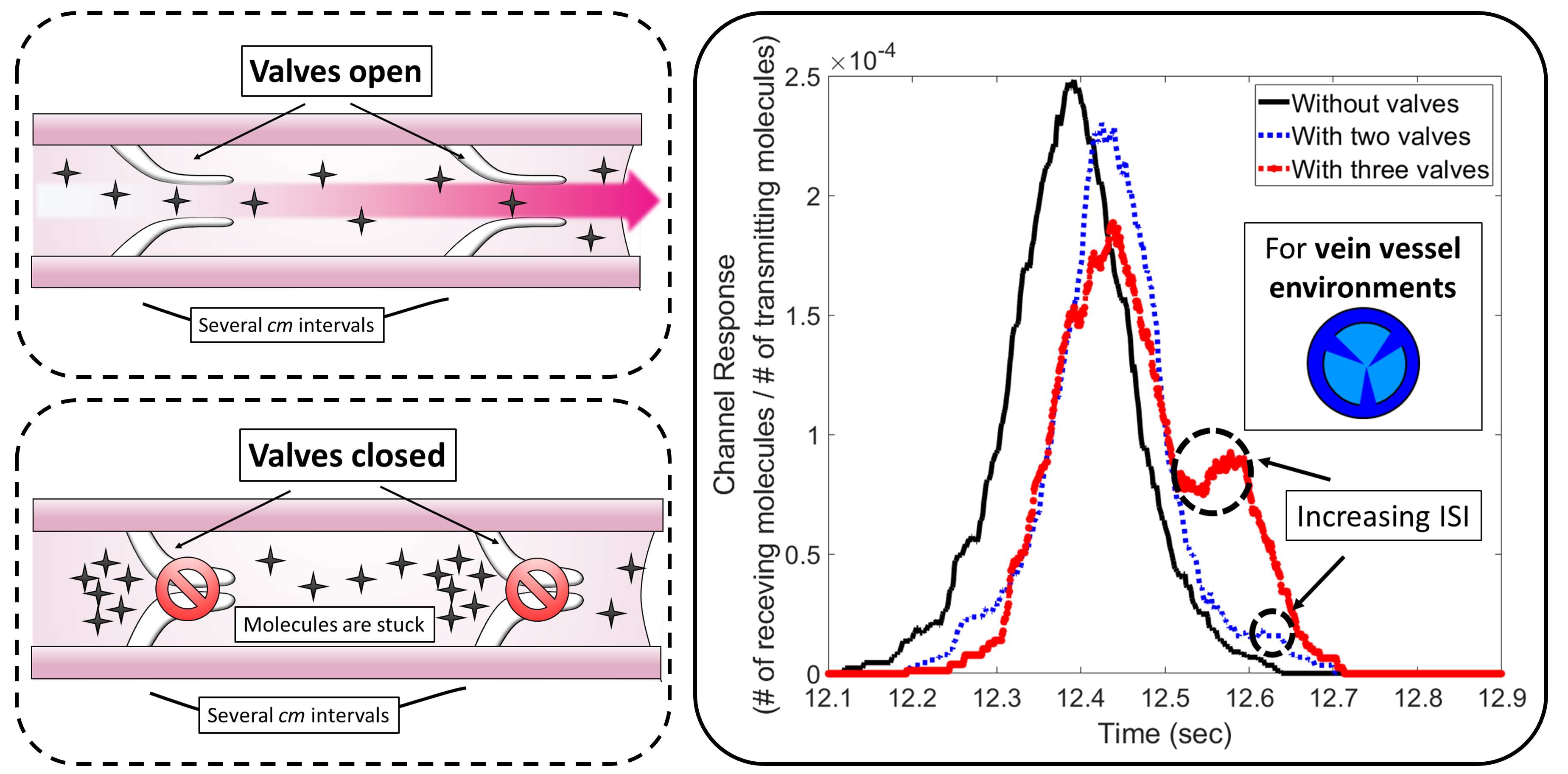}
		\label{fig_blood_valves}}
	\caption{Characteristics of blood vessels (capillaries and veins), (a) illustration of capillary blood vessels (left) and channel responses with/without leakage (right), (b) illustration of valves operation in vein blood vessels (left) and channel responses with/without values (right).}
	\label{fig_vessel_characteristics}
\end{figure*}

\subsection{Boundary Condition}
Blood vessels have different types of vessel walls based on their role and location. Furthermore, the vein has a special valve in its vessel. The artery generally has thick walls that render vessels durable against high pressure, preventing leakage. Therefore, the artery can be modeled by a cylindrical duct with a full reflection boundary if the artery does not have wall malfunctions.

On the other hand, capillaries have very thin walls for the exchange of various nutrients. A permeable boundary can cause leakage of messenger molecules. The leakage rate varies according to the size of the molecules and their interaction with the walls. Fig.~\ref{fig_leakage} shows the capillary and channel response influenced by leakage on the surface. The authors in~\cite{Arjmandi2021} introduced a cylindrical duct channel model with arbitrary boundary conditions. Based on a previous study, we can model the capillary channel model if we know the leakage rate of messenger molecules through the capillary walls.

The vein has a moderate wall thickness. However, the natural valves are located with a spacing of several centimeters to enable flow, which helps molecular movement. This makes it difficult to analyze the channel model when molecules pass through many valves. The transmission signal may be damaged by valves that open and close periodically. To the best of our knowledge, there is no proper boundary condition to help build a channel model with natural valves. Fig.~\ref{fig_blood_valves} illustrates that the valves in the vein and the operation of valves influence the channel response.

\subsection{Molecules Types}
The transformation of molecules should also be considered for safety and channel modeling. In other words, appropriate molecules should be utilized as messengers to make the health monitoring system safe. However, the receiver may not detect the transformed molecules, thus influencing the channel model.

Many natural molecules, other than messengers, are present in blood vessels. Therefore, messenger molecules may be transformed by chemical reactions with other molecules. We describe three main transformations: isomer, degradation, and synthesis.

First, isomer molecules have different shapes but with identical molecular formulas. Isomers have different chemical characteristics that may be used to indicate different information. In ~\cite{Kim2013}, the authors proposed modulation using the isomer group of glucose. If each isomer has similar physical characteristics, such as the diffusion coefficient, the channel model can be conveniently developed.

Second, the degradation and synthesis of molecules influence the molecular communication channel model. Chemical reactions in blood vessels may disassemble or assemble messenger molecules, while molecules convey information. The transmitted signal is lost if the receiver cannot detect the byproducts of degradation and synthesis. Subsequently, the channel impulse tail becomes shorter~\cite{Jamali2019}. These characteristics can be used to mitigate the ISI and increase the communication reliability if the byproducts are safe for the body. Previously, the authors in~\cite{Zoofaghari2019} analyzed a cylindrical duct channel model by applying a first-order chemical reaction. A channel model can be built when the communication system utilizes changeable molecules by choosing the proper effect. It is worth trying the machine-learning method when the environment has intractable chemical reactions.

From a safety point of view, utilizing molecules that do not have any chemical reactions in the body and eliminated naturally is better if their reliability and energy efficiency meet the requirements for health-monitoring systems. The molecules used in medical checkups are candidates for stable messenger molecules.

\subsection{Modulation \& Detection}
Previous studies on molecular communication have proposed several modulation methods such as pulse position modulation (PPM), concentration shift keying (CSK), and molecule-type shift keying (MoSK). Under the condition that the receiver detects any molecule type, MoSK modulation can be considered suitable in terms of reliability~\cite{Gao2021, Tang2021}. However, to support MoSK, the receiver becomes too complex to implant in the inner body, which is not desirable for health monitoring systems. Thus, alternatives such as PPM and CSK can be considered more suitable for eHealth applications. However, while these schemes have been widely studied for free-space channels, there is a need to properly elucidate their characteristics for blood vessel channels.

In~\cite{Farsad2016}, many versions of both modulations were introduced. The PPM modulates information by transmitting a signal at a specific time according to the modulation policy. The modulation shows advantages in energy consumption as it uses the same amount of molecules for all signals. Furthermore, sending a small number of molecules in a single shot is sufficient for decent reception performance. However, it requires a long symbol duration for higher modulation levels, leading to lower transmission speeds. Nevertheless, PPM is an acceptable candidate for our proposed system, given that it prioritizes energy efficiency over the transmission speed.

On the other hand, CSK modulates information by emitting varying quantities of molecules with respect to the designated levels. This method is analogous to amplitude modulation in RF communications. CSK modulation is very simple to operate on the transmitter. However, it may cause a high ISI if the proper level of the molecule count is not configured. The molecular communication channel impulse generated by counting molecules generally has a long tail and causes a high ISI. The flow assistance in the environment mitigates the longtail characteristic, but using coding and error correction can improve even more. Thus, CSK modulation has practical advantages owing to its low complexity.

The above modulations have generally been studied in free-space channel models. Therefore, channel characteristics such as leakage on the surface of the capillary should be considered. In the simulation with capillary environments, leakage on the surface decreases the amplitude of the desired signal, but the receiving time is not affected as much as the amplitude. Based on this, appropriate techniques should be developed utilizing the previous method. Also, we can consider utilizing hybrid modulation method in~\cite{Gursoy2021}.

Optimal detection was developed to pair each modulation. Maximum likelihood (ML) and maximum a-posteriori (MAP) are proposed for the best performance. However, it is difficult to adopt the above detection methods directly because of their complexity. A nanomachine may utilize various thresholding methods in a realistic view because it merely detects the transmitting sequence based on whether it is over the threshold. Channel characteristics are typically used to set the threshold. For example, the adaptive thresholding method reflects the ISI and interlink interference (ILI) in the threshold~\cite{Koo2021}. The system may require a threshold that differs according to the blood vessel in which it is installed.

We note that the vein channel requires the incorporation of ISI mitigation methods owing to the existence of valves. The valves help blood travel well, even at low blood pressure. However, it can disturb molecular communications in veins by causing ISI. Therefore, we should consider a detection method to mitigate the ISI. For example, a threshold can be configured to address the ISI caused by the tail part of the channel response. The channel response can show different schemes by changing the period and intervals of the valves being closed and open. Considering the valve operations, the vein channel model may be too complex to develop properly. Therefore, exploiting machine-learning techniques for this purpose is a plausible option for developing an integrated vein channel model.

\subsection{Coding \& Error Correction}

Coding and error correction techniques are necessary to achieve high reliability. However, nano-machines in the body may lack sufficient energy to carry out the complex methods used in traditional RF communications. Coding and error corrections for molecular communication were introduced in~\cite{Farsad2016, Dhayabaran2021}. Coding methods based on hamming and convolution codes have been proposed, such as RF communication. The coding methods were satisfactory in terms of their performance. However, the ISI-free or mitigating code is more acceptable for making the system more reliable.

In~\cite{Kislal2020}, ISI-mitigating channel codes were proposed for molecular communication. The codes use channel information and exhibit a lower bit error rate (BER). In the binary concentration shift keying (BCSK) modulation scheme, successive bit- 1s cause severe ISI. Therefore, the rule of the codebook was set to avoid consecutive bit-1s. The study assumes a free-space diffusion channel, but we can utilize it by replacing it with our system channel. Furthermore, the coding method may perform better because the flow-assistance channel generally has a lower ISI.

\subsection{Relay System}

The molecular communication system requires a relay system, although the blood vessel makes a flow-assistance~\cite{Wang2020}. Just one transceiver pair is hard to convey signals from the sensor that measured the biological signal to the outer body machine. The reliable and energy-efficient relay system should be considered for the health-monitoring system.

The neural system has a relay for transmitting the signal to the whole body. Furthermore, messenger molecules are reused at the synapse that is an interval between front and back neurons. In the molecular communication view, the relay with reusing molecules system was introduced in~\cite{Guo2018}.

The relay system will play an important role in the vein channel or the capillary channel. First, the vein communication system requires the relay system because of the valves in the vein. The natural valves are located by several-cm spacing for making flow which helps molecular movement. However, it makes it hard to analyze the channel model when molecules pass many valves. The transmission signal may get damage by valves that open and close periodically, as shown in Fig.~\ref{fig_blood_valves}. We propose considering that distribute transceiver pairs according to the distribution of valves. It may increase the reliability of communication, and we expect better performance if the relay period matches the period of valves. 

The capillary channel requires the system for another reason. The capillary has the lowest flow speed, and the leakage may happen at the capillary surface, as shown in Fig.~\ref{fig_leakage}. Also, it has the smallest cross-section size of the duct that makes limitations on carrying energy in nano-machines. The characteristics mentioned above make it hard to transmit molecular signals further, which can be mitigated by the relay system. 

\section{Prototypes Design}
\label{testbed}

\begin{figure*}[!t]
	\centering
	\subfloat[Meso-scale vessel MIMO prototype]{\includegraphics[width=1.5\columnwidth,keepaspectratio]{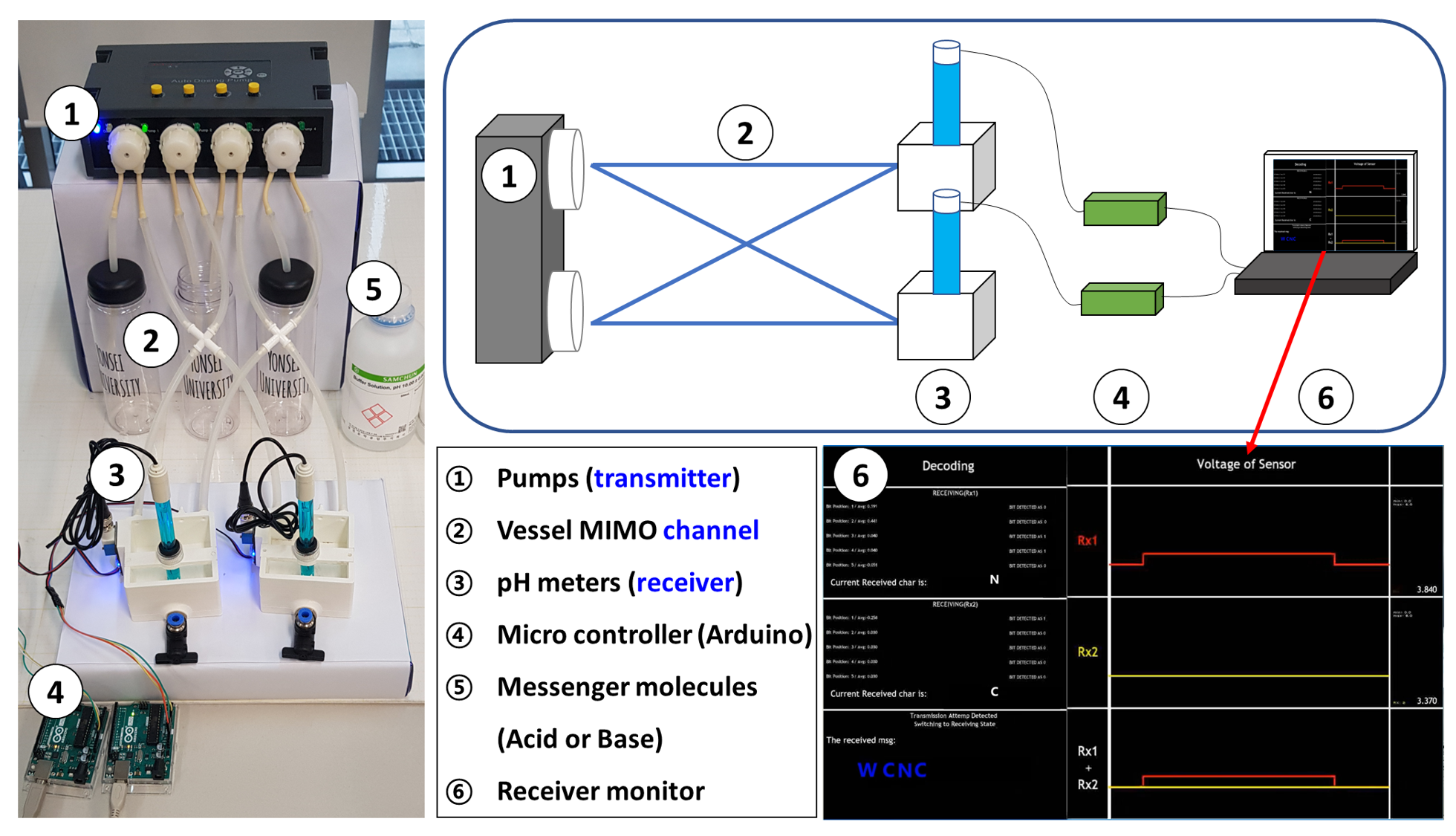}
		\label{fig_macro_prototype}}
	\hfill
	\subfloat[Nano-scale human implantable vessel prototype]{\includegraphics[width=1.5\columnwidth,keepaspectratio]{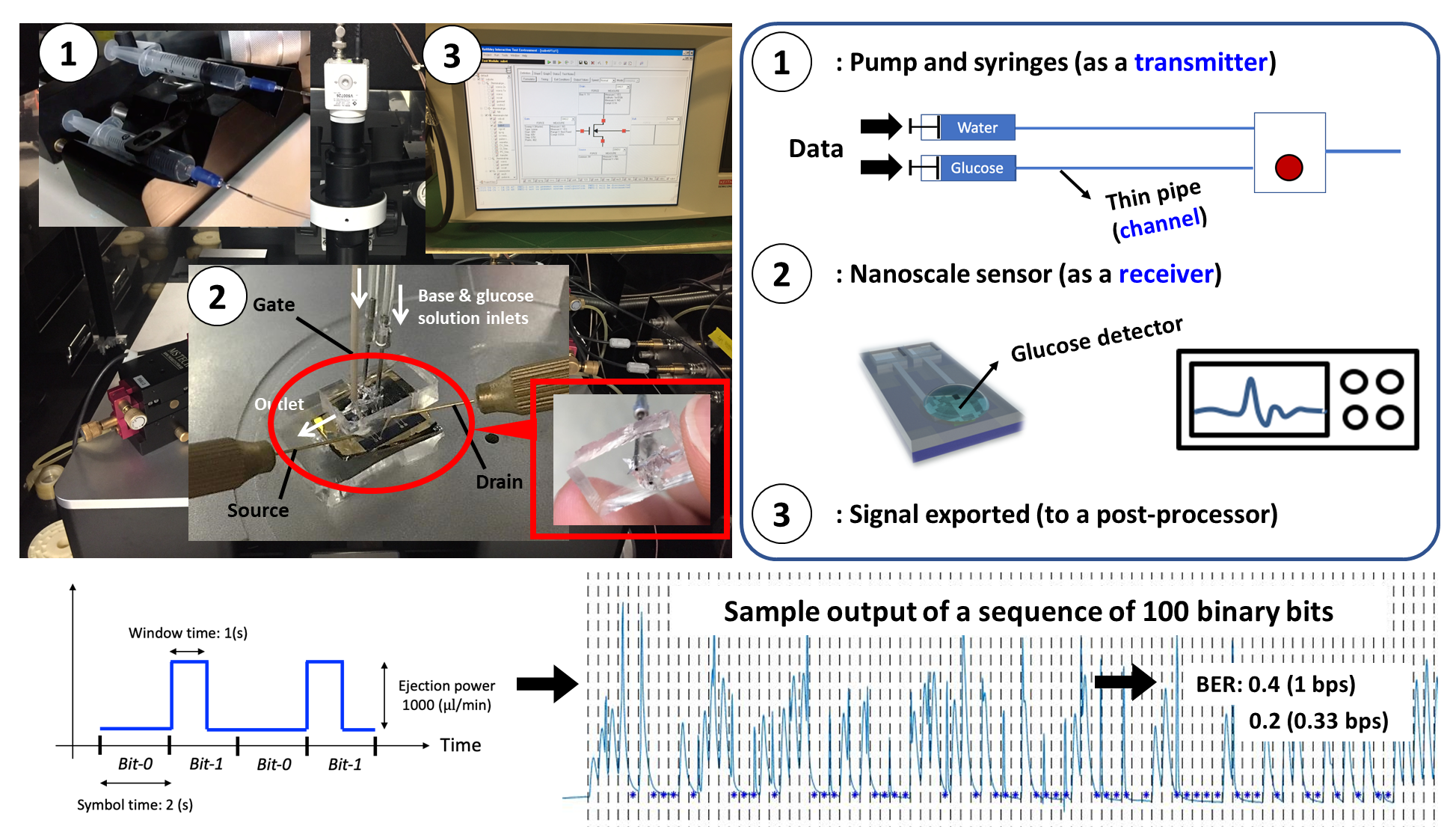}\label{fig_nano_prototype}}
	\caption{Prototypes for blood vessel networks.}
	\label{fig_prototypes}
\end{figure*}

Several studies have introduced testbeds for molecular communications to show practical possibilities. Each testbed was developed for free-space, vast water, and tube environments. Each testbed also uses different forces such as compressed air~\cite{Lee2015, Lee2015a, Koo2016, Koo2021}, liquid flow~\cite{Lemal2019, Al-Helali2020, Lee2020, Koo2020, Khaloopour2020}, magnetic force~\cite{Wicke2019, Wicke2022}, and gravity~\cite{Guo2020} to transmit molecules. Here, we introduce meso- and nano-scale testbed to verify the feasibility of molecular communication in vessels.

\subsection{Vessel MIMO Prototype}
In~\cite{Lee2020}, we introduced a meso-scale in-vessel molecular MIMO communication testbed, as shown in Fig.~\ref{fig_macro_prototype}. The transmitter was made by four pumps that make flow in the tube channel and emitted messenger molecules. The system utilizes bases and acids as messenger molecules, and each pump indeed can use different molecules. The emitted molecules pass through a 2 $\times$ 2 MIMO channel composed of junction parts and silicon tubes. The receiver measures the change of pH value in the receiver tank and converted it to an electrical signal. Finally, the computer displays the detection results on the screen using the conveyed electrical signal. The system is operated by CSK modulation, International Telegraph Alphabet no.~2 (ITA2) coding, adaptive threshold detection, and spatial multiplexing. Microcontroller Arduino controls both the transmitter and the receiver. Therefore, the testbed can utilize various modulations, coding methods, and detections. Furthermore, the testbed can compose various multiple-branches channels using junction parts made by a 3D printer.

\subsection{Human Implantable Nano Prototype}
We also implemented a nano-scale system that can communicate via molecules, as shown in Fig.~\ref{fig_nano_prototype}~\cite{Koo2020}. While the entire communication model still requires enhancements to be applied in practice, the remarkable point of the work is that the nano-scale sensor we designed is specialized for molecular communication tasks. The sensor is small enough to be implanted near the human skin while not being harmful; thus, it is suitable for being a molecular communication receiver. It can notice a sensitive amount of glucose molecules using indium gallium zinc oxide (IGZO) enzymes and report the variations of detected concentrations so that the information can be conveyed and converted. Also, the sensor has low manufacturing costs so that it can be easily replaceable, and the fact compensates demerit of a short lifetime the device has. The remaining part of the system is to demonstrate and verify the communication feasibility. Our results show that the data rate of 2 bps is achievable with a reliable error rate. We conclude with a positive message that the bottleneck is not the channel uncertainty but the mechanical units so that the nano-molecular communication system can achieve even better performances in future works.

\section{Open Challenges and Concluding Remarks}
\label{conclusion}
In this work, we investigated the system design and reported prototypes for the Internet of Bio-Nano Things (IoBNT) that can be utilized in eHealth applications. Bio-nano things are connected through molecular communications in blood vessels. We have shown that the blood vessel channels differ considerably from the free-space channel, and utilizing cylindrical duct channel models is appropriate for molecular communications. We have also proposed molecular communication system designs that guarantee high reliability and low energy consumption: building architectures tailored for the blood vessel type. We empirically showed that the channel environments of blood vessels introduce a novel set of challenges compared to free-space channels and suggest the utilization of cylindrical duct channel models as suitable candidates for molecular communication. Furthermore, we confirmed the feasibility of the proposed IoT systems using nanoscale molecular communication prototype implementations. 

Despite being in its preliminary stages, we see a number of interesting research challenges to be addressed, including a deeper understanding of biological channel environments, new components, and system architectures for supporting effective modulation, detection and coding. These results will lead to new applications and prototype implementations that can further enhance the practical applicability of IoBNT systems. We expect that this work will provide insight into the system design of IoBNT with molecular communication.

\section*{Acknowledgment}
This work was supported by the National Research Foundation of Korea (NRF) Grant through the Ministry of Science and ICT (MSIT), Korea Government (NRF-2020R1A2C4001941, 2022R1A5A1027646).


\bibliographystyle{IEEEtran}
\bibliography{IEEEabrv,JCN_Collection_Lee}

\begin{thebibliography}{10}
\providecommand{\url}[1]{#1}
\csname url@samestyle\endcsname
\providecommand{\newblock}{\relax}
\providecommand{\bibinfo}[2]{#2}
\providecommand{\BIBentrySTDinterwordspacing}{\spaceskip=0pt\relax}
\providecommand{\BIBentryALTinterwordstretchfactor}{4}
\providecommand{\BIBentryALTinterwordspacing}{\spaceskip=\fontdimen2\font plus
\BIBentryALTinterwordstretchfactor\fontdimen3\font minus
  \fontdimen4\font\relax}
\providecommand{\BIBforeignlanguage}[2]{{%
\expandafter\ifx\csname l@#1\endcsname\relax
\typeout{** WARNING: IEEEtran.bst: No hyphenation pattern has been}%
\typeout{** loaded for the language `#1'. Using the pattern for}%
\typeout{** the default language instead.}%
\else
\language=\csname l@#1\endcsname
\fi
#2}}
\providecommand{\BIBdecl}{\relax}
\BIBdecl

\bibitem{Pramanik2020}
P.~K.~D. Pramanik, A.~Solanki, A.~Debnath, A.~Nayyar, S.~El-Sappagh, and K.~S.
  Kwak, ``{Advancing Modern Healthcare with Nanotechnology, Nanobiosensors, and
  Internet of Nano Things: Taxonomies, Applications, Architecture, and
  Challenges},'' \emph{{IEEE} Access}, vol.~8, pp. 65\,230--65\,266, 2020.

\bibitem{Zafar2021}
S.~Zafar, M.~Nazir, T.~Bakhshi, H.~{Ali Khattak}, S.~Khan, M.~Bilal, K.~K.~R.
  Choo, K.~S. Kwak, and A.~Sabah, ``{A Systematic Review of Bio-Cyber Interface
  Technologies and Security Issues for Internet of Bio-Nano Things},''
  \emph{{IEEE} Access}, vol.~9, pp. 93\,529--93\,566, 2021.

\bibitem{Akyildiz2015}
I.~F. Akyildiz, M.~Pierobon, S.~Balasubramaniam, and Y.~Koucheryavy, ``{The
  Internet of Bio-Nano Things},'' \emph{{IEEE} Commun. Mag.}, vol.~53, no.~3,
  pp. 32--40, Mar. 2015.

\bibitem{Akyildiz2019}
I.~F. Akyildiz, M.~Pierobon, and S.~Balasubramaniam, ``{Moving Forward with
  Molecular Communication: From Theory to Human Health Applications},''
  \emph{Proc. {IEEE}}, vol. 107, no.~5, pp. 858--865, 2019.

\bibitem{Guo2016}
W.~Guo, T.~Asyhari, N.~Farsad, H.~B. Yilmaz, B.~Li, A.~Eckford, and C.-B. Chae,
  ``{Molecular Communications: Channel Model and Physical Layer Techniques},''
  \emph{{IEEE} Wireless Commun.}, vol.~23, no.~4, pp. 120--127, Aug. 2016.

\bibitem{Farsad2016}
N.~Farsad, H.~B. Yilmaz, A.~Eckford, C.-B. Chae, and W.~Guo, ``{A Comprehensive
  Survey of Recent Advancements in Molecular Communication},'' \emph{{IEEE}
  Commun. Surveys Tuts.}, vol.~18, no.~3, pp. 1887--1919, Feb. 2016.

\bibitem{Jamali2019}
V.~Jamali, A.~Ahmadzadeh, W.~Wicke, A.~Noel, and R.~Schober, ``{Channel
  Modeling for Diffusive Molecular Communication-A Tutorial Review},''
  \emph{Proc. {IEEE}}, vol. 107, no.~7, pp. 1256--1301, Jul. 2019.

\bibitem{Lo2019}
Y.~F. Lo, C.~H. Lee, P.~C. Chou, and P.~C. Yeh, ``{Modeling Molecular
  Communications in Tubes with Poiseuille Flow and Robin Boundary Condition},''
  \emph{{IEEE} Commun. Lett.}, vol.~23, no.~8, pp. 1314--1318, 2019.

\bibitem{Schafer2021}
M.~Sch{\"{a}}fer, W.~Wicke, L.~Brand, R.~Rabenstein, and R.~Schober,
  ``{Transfer Function Models for Cylindrical MC Channels with Diffusion and
  Laminar Flow},'' \emph{{IEEE} Trans. Mol. Bio. Multi-Scale Commun.}, vol.~7,
  no.~4, pp. 271--287, 2021.

\bibitem{Dhok2022}
S.~Dhok, L.~Chouhan, A.~Noel, and P.~K. Sharma, ``{Cooperative Molecular
  Communication in Drift-Induced Diffusive Cylindrical Channel},'' \emph{{IEEE}
  Trans. Mol. Bio. Multi-Scale Commun.}, vol.~8, no.~1, pp. 44--55, 2022.

\bibitem{Dhok2022a}
S.~Dhok, P.~Peshwe, and P.~K. Sharma, ``{Cognitive Molecular Communication in
  Cylindrical Anomalous-Diffusive Channel},'' \emph{{IEEE} Trans. Mol. Bio.
  Multi-Scale Commun.}, vol.~8, no.~3, pp. 158--168, 2022.

\bibitem{Varshney2018}
N.~Varshney, A.~Patel, Y.~Deng, W.~Haselmayr, P.~K. Varshney, and
  A.~Nallanathan, ``{Abnormality Detection Inside Blood Vessels with Mobile
  Nanomachines},'' \emph{{IEEE} Trans. Mol. Bio. Multi-Scale Commun.}, vol.~4,
  no.~3, pp. 189--194, 2018.

\bibitem{Zoofaghari2019}
M.~Zoofaghari and H.~Arjmandi, ``{Diffusive Molecular Communication in
  Biological Cylindrical Environment},'' \emph{{IEEE} Trans. NanoBiosci.},
  vol.~18, no.~1, pp. 74--83, Jan. 2019.

\bibitem{Arjmandi2021}
H.~Arjmandi, M.~Zoofaghari, S.~V. Rouzegar, M.~Veletic, and I.~Balasingham,
  ``{On Mathematical Analysis of Active Drug Transport Coupled with
  Flow-Induced Diffusion in Blood Vessels},'' \emph{{IEEE} Trans. NanoBiosci.},
  vol.~20, no.~1, pp. 105--115, Jan. 2021.

\bibitem{Lee2020}
C.~Lee, B.-H. Koo, and C.-B. Chae, ``{Demo: In-Vessel Molecular MIMO
  Communications},'' in \emph{Proc. IEEE Wireless Commun. and Netw. Conf.
  Workshops (WCNCW)}, 2020, pp. 6--7.

\bibitem{Koo2020}
B.-H. Koo, H.~J. Kim, J.~Y. Kwon, and C.-B. Chae, ``{Deep Learning-based Human
  Implantable Nano Molecular Communications},'' in \emph{Proc. IEEE Int. Conf.
  on Commun. (ICC)}, Jun. 2020.

\bibitem{Campbell2016}
L.~A. Urry, M.~L. Cain, S.~A. Wasserman, P.~V. Minorsky, J.~B. Reece, and N.~A.
  Campbell, \emph{Campbell Biology, 11th ed.}\hskip 1em plus 0.5em minus
  0.4em\relax London, England: Pearson, 2016.

\bibitem{Abbasi2017}
N.~A. Abbasi and O.~B. Akan, ``{An Information Theoretical Analysis of Human
  Insulin-Glucose System Toward the Internet of Bio-Nano Things},''
  \emph{{IEEE} Trans. NanoBiosci.}, vol.~16, no.~8, pp. 783--791, 2017.

\bibitem{Akyildiz2020}
I.~F. Akyildiz, M.~Ghovanloo, U.~Guler, T.~Ozkaya-Ahmadov, A.~F. Sarioglu, and
  B.~D. Unluturk, ``{PANACEA: An Internet of Bio-NanoThings Application for
  Early Detection and Mitigation of Infectious Diseases},'' \emph{{IEEE}
  Access}, vol.~8, pp. 140\,512--140\,523, 2020.

\bibitem{Misra2020}
S.~Misra, S.~Pal, S.~Kaneriya, S.~Tanwar, N.~Kumar, and J.~P. {Joel Rodrigues},
  ``{Population Dynamics of Biosensors for Nano-therapeutic Applications in
  Internet of Bio-Nano Things},'' in \emph{Proc. IEEE Int. Conf. on Commun.
  (ICC)}, Jun. 2020.

\bibitem{El-Fatyany2020}
A.~El-Fatyany, H.~Wang, and S.~M. {Abd El-atty}, ``{On Mixing Reservoir
  Targeted Drug Delivery Modeling-based Internet of Bio-NanoThings},''
  \emph{Wireless Netw.}, vol.~26, no.~5, pp. 3701--3713, 2020.

\bibitem{Dissanayake2021}
M.~B. Dissanayake and N.~Ekanayake, ``{On the Exact Performance Analysis of
  Molecular Communication via Diffusion for Internet of Bio-Nano Things},''
  \emph{{IEEE} Trans. NanoBiosci.}, vol.~20, no.~3, pp. 291--295, 2021.

\bibitem{Al-Zubi2022}
M.~M. Al-Zubi, A.~S. Mohan, P.~Plapper, and S.~H. Ling, ``{Intra-Body Molecular
  Communication via Blood-Tissue Barrier for Internet of Bio-Nano Things},''
  \emph{{IEEE} Internet Things J.}, no.~c, pp. 1--9, 2022.

\bibitem{Juwono2021}
F.~H. Juwono, R.~Reine, W.~K. Wong, Z.~A. Sim, and L.~Gopal, ``{Envisioning 6G
  Molecular Communication for IoBNT Diagnostic Systems},'' in \emph{Proc. Int.
  Conf. on Green Energy, Computing and Sustainable Technology (GECOST)}, 2021,
  pp. 1--5.

\bibitem{Swaminathan2017}
M.~Swaminathan, A.~Vizziello, D.~Duong, P.~Savazzi, and K.~R. Chowdhury,
  ``{Beamforming In The Body: Energy-Efficient and Collision-Free Communication
  for Implants},'' in \emph{Proc. IEEE Int. Conf. on Comput. Commun.
  (INFOCOM)}, 2017, pp. 1--9.

\bibitem{Chude-Okonkwo2016}
U.~A. Chude-Okonkwo, R.~Malekian, and B.~T. Maharaj, ``{Biologically Inspired
  Bio-Cyber Interface Architecture and Model for Internet of Bio-Nanothings
  Applications},'' \emph{{IEEE} Trans. Commun.}, vol.~64, no.~8, pp.
  3444--3455, 2016.

\bibitem{Yang2020}
K.~Yang, D.~Bi, Y.~Deng, R.~Zhang, M.~M. {Ur Rahman}, N.~A. Ali, M.~A. Imran,
  J.~M. Jornet, Q.~H. Abbasi, and A.~Alomainy, ``{A Comprehensive Survey on
  Hybrid Communication in Context of Molecular Communication and Terahertz
  Communication for Body-Centric Nanonetworks},'' \emph{{IEEE} Trans. Mol. Bio.
  Multi-Scale Commun.}, vol.~6, no.~2, pp. 107--133, 2020.

\bibitem{Kim2013}
N.-R. Kim and C.-B. Chae, ``{Novel Modulation Techniques using Isomers as
  Messenger Molecules for Nano Communication Networks via Diffusion},''
  \emph{{IEEE} J. Sel. Areas Commun.}, vol.~31, no.~12, pp. 847--856, 2013.

\bibitem{Gao2021}
W.~Gao, T.~Mak, and L.-l. Yang, ``{Molecular Type Spread Molecular Shift Keying
  for Multiple-Access Diffusive Molecular Communications},'' \emph{{IEEE}
  Trans. Mol. Bio. Multi-Scale Commun.}, vol.~7, no.~1, pp. 51--63, 2021.

\bibitem{Tang2021}
Y.~Tang, Y.~Huang, C.-B. Chae, W.~Duan, M.~Wen, and L.~L. Yang,
  ``{Molecular-Type Permutation Shift Keying in Molecular MIMO Communications
  for IoBNT},'' \emph{{IEEE} Internet Things J.}, vol.~8, no.~21, pp.
  16\,023--16\,034, 2021.

\bibitem{Gursoy2021}
M.~C. Gursoy, D.~Seo, and U.~Mitra, ``{A Concentration-Time Hybrid Modulation
  Scheme for Molecular Communications},'' \emph{{IEEE} Trans. Mol. Bio.
  Multi-Scale Commun.}, vol.~7, no.~4, pp. 288--299, 2021.

\bibitem{Koo2021}
B.-H. Koo, C.~Lee, A.~E. Pusane, T.~Tugcu, and C.-B. Chae, ``{MIMO Operations
  in Molecular Communications: Theory, Prototypes, and Open Challenges},''
  \emph{{IEEE} Commun. Mag.}, vol.~59, no.~9, pp. 98--104, 2021.

\bibitem{Dhayabaran2021}
B.~Dhayabaran, G.~T. Raja, and M.~Magarini, ``{Low Complex Receiver Design for
  Modified Inverse Source Coded Diffusion-Based Molecular Communication
  Systems},'' \emph{{IEEE} Trans. Mol. Bio. Multi-Scale Commun.}, vol.~7,
  no.~4, pp. 239--252, 2021.

\bibitem{Kislal2020}
A.~O. Kislal, B.~C. Akdeniz, C.~Lee, A.~E. Pusane, T.~Tugcu, and C.-B. Chae,
  ``{ISI-Mitigating Channel Codes for Molecular Communication Via Diffusion},''
  \emph{{IEEE} Access}, vol.~8, pp. 24\,588--24\,599, Jan. 2020.

\bibitem{Wang2020}
J.~Wang, M.~Peng, Y.~Liu, X.~Liu, and M.~Daneshmand, ``{Performance Analysis of
  Signal Detection for Amplify-and-Forward Relay in Diffusion-Based Molecular
  Communication Systems},'' \emph{{IEEE} Internet Things J.}, vol.~7, no.~2,
  pp. 1401--1412, 2020.

\bibitem{Guo2018}
W.~Guo, Y.~Deng, H.~B. Yilmaz, N.~Farsad, M.~Elkashlan, A.~Eckford,
  A.~Nallanathan, and C.-B. Chae, ``{SMIET: Simultaneous Molecular Information
  and Energy Transfer},'' \emph{{IEEE} Wireless Commun.}, vol.~25, no.~1, pp.
  106--113, Feb. 2018.

\bibitem{Lee2015}
C.~Lee, B.~Koo, N.-R. Kim, H.~B. Yilmaz, N.~Farsad, A.~Eckford, and C.-B. Chae,
  ``{Molecular MIMO Communication Link},'' in \emph{Proc. IEEE Int. Conf. on
  Comput. Commun. (INFOCOM)}.\hskip 1em plus 0.5em minus 0.4em\relax IEEE,
  2015, pp. 13--14.

\bibitem{Lee2015a}
C.~Lee, B.~Koo, N.~R. Kim, H.~B. Yilmaz, N.~Farsard, A.~Eckford, and C.~B.
  Chae, ``{Demo: Molecular MIMO with drift},'' in \emph{Proc. the 21st Annual
  Int. Conf. on Mobile Comput. and Netw. (MOBICOM)}, 2015, pp. 201--203.

\bibitem{Koo2016}
B.-H. Koo, C.~Lee, H.~B. Yilmaz, N.~Farsad, A.~Eckford, and C.-B. Chae,
  ``{Molecular MIMO: From Theory to Prototype},'' \emph{{IEEE} J. Sel. Areas
  Commun.}, vol.~34, no.~3, pp. 600--614, Mar. 2016.

\bibitem{Lemal2019}
P.~Lemal, A.~Petri-Fink, and S.~Balog, ``{Nanoparticles and Taylor Dispersion
  as a Linear Time-Invariant System},'' \emph{Analytical Chemistry}, vol.~91,
  no.~2, pp. 1217--1221, 2019.

\bibitem{Al-Helali2020}
A.~Al-Helali, B.~Liang, and N.~Nasser, ``{Novel Molecular Signaling Method and
  System for Molecular Communication in Human Body},'' \emph{{IEEE} Access},
  vol.~8, pp. 119\,361--119\,375, Jun. 2020.

\bibitem{Khaloopour2020}
L.~Khaloopour, M.~Nasiri-Kenari, S.~V. Rouzegar, A.~Azizi, A.~Hosseinian,
  M.~Farahnak-Ghazani, N.~Bagheri, M.~Mirmohseni, H.~Arjmandi, and R.~Mosayebi,
  ``{An Experimental Platform for Macro-Scale Fluidic Medium Molecular
  Communication},'' \emph{{IEEE} Trans. Mol. Bio. Multi-Scale Commun.}, vol.~5,
  no.~3, pp. 163--175, 2020.

\bibitem{Wicke2019}
W.~Wicke, A.~Ahmadzadeh, V.~Jamali, H.~Unterweger, C.~Alexiou, and R.~Schober,
  ``{Magnetic Nanoparticle-Based Molecular Communication in Microfluidic
  Environments},'' \emph{{IEEE} Trans. NanoBiosci.}, vol.~18, no.~2, pp.
  156--169, 2019.

\bibitem{Wicke2022}
W.~Wicke, H.~Unterweger, J.~Kirchner, L.~Brand, A.~Ahmadzadeh, D.~Ahmed,
  V.~Jamali, C.~Alexiou, G.~Fischer, and R.~Schober, ``{Experimental System for
  Molecular Communication in Pipe Flow With Magnetic Nanoparticles},''
  \emph{{IEEE} Trans. Mol. Bio. Multi-Scale Commun.}, vol.~8, no.~2, pp.
  56--71, 2022.

\bibitem{Guo2020}
W.~Guo, I.~Atthanayake, and P.~Thomas, ``{Vertical Underwater Molecular
  Communications via Buoyancy: Gaussian Velocity Distribution of Signal},'' in
  \emph{Proc. IEEE Int. Conf. on Commun. (ICC)}, 2020.

\end{thebibliography}

\newpage

\begin{IEEEbiography}[{\includegraphics[width=1in,height=1.25in,clip,keepaspectratio]{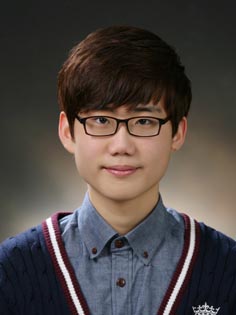}}]
	{Changmin Lee}
	(S'15)  received the B.S. degree
	from the School of Integrated Technology, Yonsei
	University, Seoul, South Korea, in 2015. He is currently
	pursuing the Ph.D. degree at the School of
	Integrated Technology, Yonsei University. He was
	the recipient of the Best Demo Award at IEEE WCNC (2020) and the Best Demo Award at IEEE
	INFOCOM (2015). His research interests include
	emerging communication technologies. 
\end{IEEEbiography}

\begin{IEEEbiography}[{\includegraphics[width=25mm,clip,keepaspectratio]{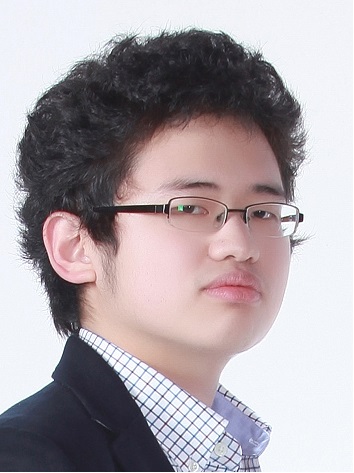}}]{Bon-Hong Koo}
	received his B.S. degree in the School of Integrated Technology from Yonsei University, Korea, in 2014. He is now with the School of Integrated Technology, at the same university and is working toward the Ph.D. degree. He was the co-recipient of the best demo awards at IEEE WCNC (2020) and IEEE INFOCOM (2015). His research interest includes emerging communication technologies.
\end{IEEEbiography}

%

\begin{IEEEbiography}[{\includegraphics[width=1in,height=1.25in,clip,keepaspectratio]{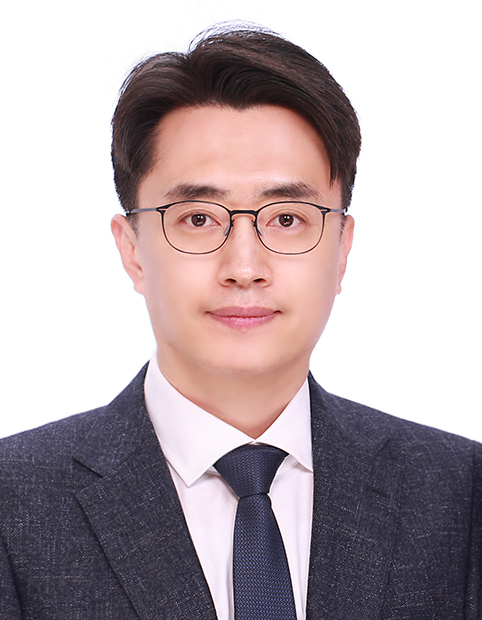}}]{Chan-Byoung~Chae}
	(S'06-M'09-SM'12-F'21)  is an Underwood Distinguished Professor in the School of Integrated Technology, Yonsei University, Korea. Before joining Yonsei University, he was with Bell Labs, Alcatel-Lucent, Murray Hill, NJ, USA from 2009 to 2011, as a Member of Technical Staff, and Harvard University, Cambridge, MA, USA from 2008 to 2009, as a Postdoctoral Research Fellow. He received his Ph.D. degree in Electrical \& Computer Engineering from The University of Texas at Austin in 2008. Prior to joining UT, he was a research engineer at the Telecommunications R\&D Center, Samsung Electronics, Suwon, Korea, from 2001 to 2005.
	
	He is now an Editor-in-Chief of the IEEE Trans. Molecular, Biological, and Multi-scale Communications and a Senior Editor of the IEEE Wireless Communications Letters. He has served/serves as an Editor for the IEEE Communications Magazine (2016-present), the IEEE Trans. on Wireless Communications (2012-2017), and the IEEE Wireless Communications Letters (2016-present). He is an IEEE ComSoc Distinguished Lecturer for the term 2020-2021 and 2022-2023.
	
	He was the recipient/co-recipient of the IEEE ICC Best Demo Award  in 2022, the IEEE WCNC Best Demo Award in 2020, the Best Young Engineer Award from the National Academy of Engineering of Korea (NAEK) in 2019, the IEEE DySPAN Best Demo Award in 2018, the IEEE/KICS Journal of Communications and Networks Best Paper Award in 2018, the IEEE INFOCOM Best Demo Award in 2015, the IEIE/IEEE Joint Award for Young IT Engineer of the Year in 2014, the KICS Haedong Young Scholar Award in 2013, the IEEE Signal Processing Magazine Best Paper Award in 2013, the IEEE ComSoc AP Outstanding Young Researcher Award in 2012, the IEEE VTS Dan. E. Noble Fellowship Award in 2008.
\end{IEEEbiography}

\begin{IEEEbiography}[{\includegraphics[width=1in,height=1.25in,clip,keepaspectratio]{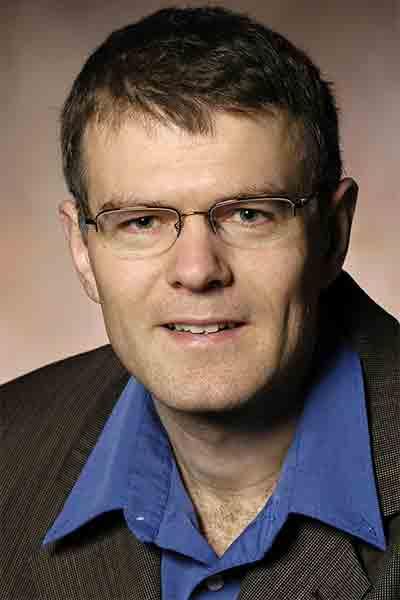}}]{Robert Schober} (S'98 - M'01 - SM'08 - F'10) received the Diplom (Univ.) and the Ph.D. degrees in electrical engineering from Friedrich-Alexander University of Erlangen-Nuremberg (FAU), Germany, in 1997 and 2000, respectively. From 2002 to 2011, he was a Professor and Canada Research Chair at the University of British Columbia (UBC), Vancouver, Canada. Since January 2012 he is an Alexander von Humboldt Professor and the Chair for Digital Communication at FAU. His research interests fall into the broad areas of Communication Theory, Wireless Communications, and Statistical Signal Processing. 
	
\end{IEEEbiography}

\end{document}